\documentstyle[12pt,aaspp4]{article}

\lefthead{Intermittency in the photosphere and corona}
\righthead{Abramenko, Yurchyshyn and Wang}

\begin{document}

\title{Intermittency in the photosphere and corona above an active region}

\author{Valentyna Abramenko}
\affil{Big Bear Solar Observatory, 40386 N. Shore Lane, Big Bear City, CA
92314}

\author{Vasyl Yurchyshyn}
\affil{Big Bear Solar Observatory, 40386 N. Shore Lane, Big Bear City, CA
92314}

\author{Haimin Wang}
\affil{Center for Solar-Terrestrial Research, New Jersey Institute of
Technology, 323 Martin Luther King Boulevard, 101 Tiernan Hall, Newark, NJ
07102}
\affil{Big Bear Solar Observatory, 40386 N. Shore Lane, Big Bear City, CA
92314}

\begin{abstract}
Recent studies undoubtedly demonstrate that the magnetic field in the
photosphere and corona is an intermittent structure, which offers new views
on the underlying physics. In particular, such problems as the existence in the
corona of localized areas with extremely strong resistivity (required to explain
magnetic reconnection of all scales) and the interchange between small and large
scales (required in study of the photosphere/corona coupling), to name a few,
can be easily captured by the concept of intermittency. This study is focused on
simultaneous time variations of intermittency properties derived in the
photosphere, chromosphere and corona. We analyzed data for NOAA AR 10930
acquired between Dec 08, 2006 12:00 UT and Dec 13, 2006 18:45 UT. Photospheric
intermittency was inferred from Hinode magnetic field measurements, while
intermittency in the transition region and corona was derived from Nobeyama 9
GHz radio polarization measurements, high cadence Hinode/XRT/Be-thin data as
well as GOES 1-8\AA~ flux. Photospheric dynamics and its possible relationship
with the intermittency variations were also analyzed by calculating the 
kinetic vorticity. For this case study we found the
following chain of events. Intermittency of the photospheric magnetic field
peaked after the specific kinetic vorticity of plasma flows in the AR reached
its maximum level (4 hour time delay). In turn, gradual increase of coronal
intermittency occurred after the peak of the photospheric intermittency. The
time delay between the peak of photospheric intermittency and the occurrence of
the first strong (X3.4) flare was approximately 1.3 days. Our analysis seems to
suggest that the enhancement of intermittency/complexity first occurs in the
photosphere and is later transported toward the corona.
\end{abstract}

\keywords{Sun: magnetic field; photosphere; corona}

\section { Introduction}

It is a widely spread view that eruptive processes of energy release in the
corona are seemingly independent of the dynamics observed in the
moderately-varying photosphere. This view is partially based on the fact that no
one-to-one correlation was observed between coronal and photospheric dynamics.
For example, numerous attempts to find any persistent {\it pre}-flare changes in
the photosphere did not lead to any solid conclusions yet. Although, recent
efforts to detect flare-related changes in the photospheric magnetic fields were
more successful (Spirock et al. 2002; Wang et al. 2004; Sudol \& Harvey 2005;
Wang 2006). More than a dozen publications by different research groups reported
that persistent abrupt changes in the photospheric magnetic flux occur in
association with X-class flares. The most plausible explanation of the observed
phenomenon seems to be flare-related changes of the inclination of field lines
rooted in the photosphere. If this is the case (future analysis of high cadence
vector magnetograms could be helpful), then we probably deal with a feedback to
the photosphere of reorganizing coronal fields.

The question whether there is any forward reaction of the photosphere toward the
corona is still open, however. There seems to be some acceptance that a {\it
statistical} relationship may exist between the conditions in the photosphere
and corona: for a large enough statistical ensemble of active regions a good
correlation has been found between photospheric magnetic parameters and coronal
phenomena (Fisher et al. 1998; Falconer et al. 2003; Schrijver et al. 2004;
Abramenko 2005a; McAteer et al. 2005; Schrijver \& Title 2005; Abramenko et al.
2006; Jing et al. 2006; Tan et al. 2007; Leka \& Barnes 2007 and references in;
 Georgoulis et al. 2007; Conlon et al. 2007). From a theoretical standpoint the
consensus seems to be reached that the ultimate source of energy for coronal
energy release is the photospheric and sub-photospheric motions of magnetized
plasma. Various mechanisms have been suggested for the energy transport toward
the corona that can and should be observationally validated (see reviews of
coronal heating mechanisms, e.g., Malara \& Velli 1994; Schrijver \& Title 2005;
Klimchuk 2006). Although, various statistical correlations do not seem to be
adequate any longer and timing characteristics are required.

Indeed, when we address issues of coupling between any two systems, a question
of vital importance immediately arises: do events that occur in one system
persistently precede or follow events in another system? If the answer is
``yes'' then what is the characteristic time delay between a pair of related
events? A shorter delay suggests a more close and intimate coupling, while long
time delays may indicate less straightforward and more complex relationship. In
particular, in such dynamical systems as magnetized turbulent plasma, the
interplay between scales may influence the delay time. In this case it is
advantageous to adopt an intermittency approach, a technique that allows
capturing interactions between various scales, in the best way possibly as of
today. In the present study, we undertook an attempt to detect the delay
intervals between key moments of the intermittency behavior in the photosphere
and corona.

\section{\bf Data Analysis}

High spatial and temporal resolution measurements of the photospheric magnetic
field and solar corona performed recently by a set of {\it Hinode} instruments
(Kosugi et al. 2007) provide us with a unique opportunity to simultaneously
estimate degree of intermittency in the photosphere and corona and to track
their variations in time.

\begin{figure}[!h] \centerline{\epsfxsize=6.0truein
\epsffile{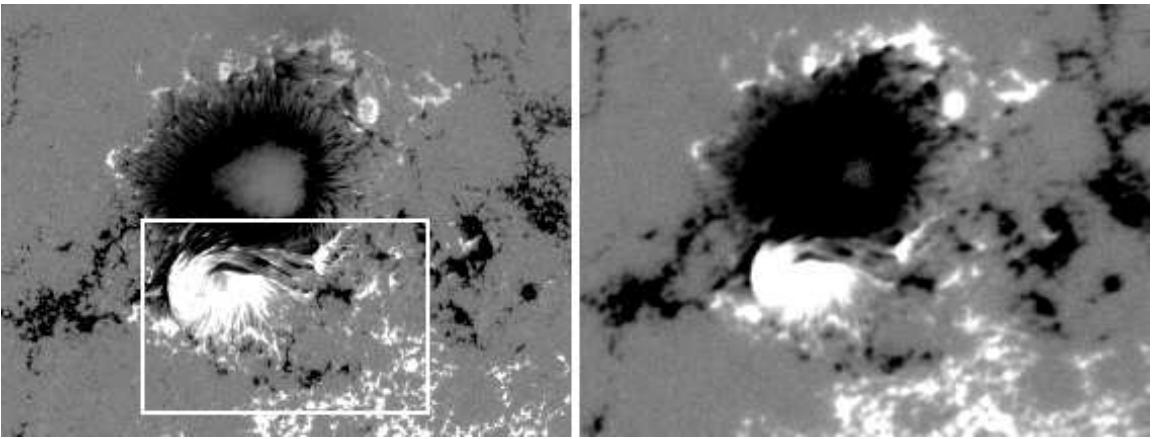}}
\caption{\sf Line-of-sight magnetograms of NOAA 10930 recorded on 2006 Dec 12 at
00:48UT by SOT-FG/Hinode ({\it left}) and SOHO/MDI HR ({\it right}). MDI/HR
(SOT/FG) magnetograms scaled between -300...300 G (DN). {\it The box } encloses
an area used for calculations of the intermittency index, $\kappa$, and the
squared kinetic vorticity, $<\omega^2>$.}
\label{Fig_1}
\end{figure}

{\it Hinode} SOT/FG instrument is designed to produce filter-based vector
magnetograms at high spatial and temporal resolution. Strictly speaking, these
images only represent measurements of the Stockes V polarization parameters at a
single wavelength so that information on the magnetic field intensity is not
available in these data. However, these data do bear information on the magnetic
structuring with an unprecedented spatial resolution of 0.16 arcsec. These
uninterrupted measurements are taken with a high time cadence of 2 minutes and
cover several days of observations of an active region. An example of a
SOT/FG/Hinode magnetogram for NOAA AR 10930 is shown in Figure \ref{Fig_1} ({\it
left}), while the right frame shows a simultaneous MDI/HR magnetogram recorded
on the same area on the Sun. As far as spatial resolution is concerned, the
advantage of the SOT/FG data is obvious.

We will analyze properties of photospheric intermittency by utilizing i) a
flatness-function technique, which relates the slope of the function to the
degree of intermittency and ii) calculations of the kinetic vorticity in the
photosphere.

Three independent data sets were used to calculate the intermittency in the
corona and chromosphere: X-ray emission records from the Hinode/XRT and GOES
instruments as well as Nobeyama 9.4 GHz polarization flux.

We analyzed active region NOAA 10930 observed by Hinode in December 2006. During
the passage across the solar disk, this active region displayed at least two
periods of enhanced activity separated by a long interval of relative quietness.
The first flaring interval lasted from Dec 4 till Dec 7, 2006, when the active
region was very close to the east limb. As evidenced from a series of MDI full
disk magnetograms, at the end of this activity period the magnetic complexity in
the active region was nearly exhausted and a new period of complexity gain was
setting in. This second activity period was accompanied by gradual emerging of a
fast-rotating sunspot of positive (N) polarity located in the close vicinity to
the main negative polarity sunspot. The emergence and rotation of the sunspot
ceased with the occurrence of two powerful X-class flares on Dec 13 and 14,
2006. The time period between Dec 8, 12:00 UT and Dec 13, 18:45 UT was chosen
to study how time variations of magnetic complexity in the photosphere and
corona are related.

\begin{figure}[!h] \centerline{\epsfxsize=5.5truein
\epsffile{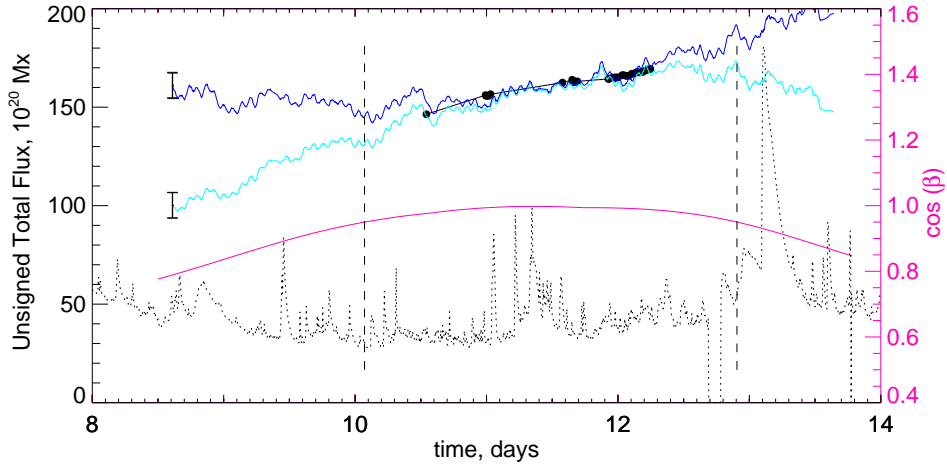}}
\caption{\sf Time variation of the unsigned magnetic flux in NOAA 10930
calculated from SOT/FG magnetograms (arbitrary units): {\it light blue - }
unsigned longitudinal flux, $\Phi_{\Vert}$; {\it dark blue - } unsigned
perpendicular to the solar surface flux, $\Phi_{\perp}$. The data points were
smoothed by 35 point box car averaging. Typical error bars are shown. {\it Pink
- } time variations of the cosine of the angular distance of the center of a
magnetogram from the center of the solar disk, $cos (\beta)$ ({\it left axis}).
Inside the interval between two vertical dashed segments, where $cos {\beta} >
0.95$, the projection effect is a minimum. {\it Dotted line} represents the GOES
1-8\AA~flux data. {\it Black circles} - longitudinal flux derived from MDI/HR
magnetograms in units of $10^{20}$ Mx ({\it left axis}).}
\label{Fig_2} 
\end{figure}

We analyzed 1718 SOT/FG $2 \times 2$ re-binned magnetograms taken with the time
cadence of 4 minutes. During the analyzed time interval the active region moved
across the solar disk from longitude of E37 to W32, so that the influence
of projection effect should be considered. We integrated longitudinal flux
density, $|B_{\Vert}|$, over the entire number, $N$, of pixels of size $\Delta
S=0.32'' \times 0.32''$ occupied by the active region. We thus obtained the
longitudinal total unsigned flux, $\Phi_{\Vert}$, which is plotted in Figure
\ref{Fig_2} with the light blue curve. The assumption that the magnetic field in
the photosphere is predominantly vertical to the solar surface offers a
possibility to estimate the magnetic flux perpendicular to the solar
surface, $\Phi_{\perp}$ (Murray 1992; Hagenaar 2001). For each magnetogram, we
calculated the cosine of the angular distance from the center of the magnetogram
to the solar disk center, $cos \beta$ (Figure \ref{Fig_2}, pink curve).
The perpendicular flux density thus can be estimated as $B_{\Vert}/cos (\beta)$
and the deprojected pixel area is $\Delta S / cos(\beta)$, which results in the
estimated magnitude of the perpendicular unsigned flux $\Phi_{\perp} =
\Phi_{\Vert}/cos^2 (\beta)$ (Figure \ref{Fig_2}, dark blue curve). Inside the
time range where $cos (\beta) > 0.95$ (i.e., between the dashed vertical line
segments in Figure \ref{Fig_2}) the perpendicular and longitudinal fluxes
differ by less than 10\%. We, therefore, accept that within this time range our
data and results are essentially free from the projection effect.

\section {\bf Measure of intermittency}

Intermittency manifests itself in both spatial (2D or 3D) and temporal (1D)
domains. In spatial domain, intermittency implies a tendency of the magnetic
field to concentrate into small-scale flux tubes of high intensity, surrounded
by extended areas of much weaker field. This tendency becomes more pronounced as
the spatial resolution of data increases. In temporal domain, intermittency is
evidenced via burst-like behavior of events. Studies of intermittency in both
spatial and temporal domains can be conducted by using the same techniques such
as the structure function approach (see, e.g., Frisch 1995).

Structure function, defined as statistical moments of the increment of a field,
is a useful tool for exploring intermittency (Stolovitzky \& Sreenivasan 1993;
Frisch 1995; Consolini et al. 1999; Abramenko et al. 2002, 2003; Abramenko
2005b; Buchlin et al. 2006; Uritsky et al. 2007).
\begin{figure}[!h] \centerline {\epsfxsize=5.0truein 
\epsffile{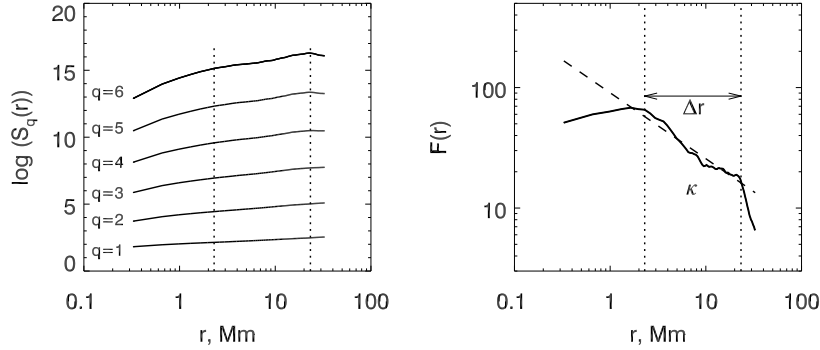}}
\caption{\sf Structure functions $S_q(r)$ (Eq. (\ref{Sq}), {\it left axis})
obtained from SOT/FG magnetogram shown in Figure \ref{Fig_1} ({\it the box}).
{\it Right :} - flatness function $F(r)$ derived from the structure functions
(Eq.(\ref{Fr})). Vertical dotted lines mark the intermittency interval, $\Delta
r$, where flatness grows according to the power law when $r$ decreases. The
interval $\Delta r$ is also marked in the left frame. The intermittency index
$\kappa$ is the slope of $F(r)$ determined within $\Delta r$: the flatness
function steepens as the magnetic field becomes more intermittent. For the
shown magnetogram, $\kappa =0.55\pm 0.05$ was determined from the linear
regression ({\it dashed line}) inside the range $\Delta r = (2.3 - 23)$ Mm.}
\label{Fig_3}
\end{figure} 

In our case the analyzed field is the line-of-sight component, $B_{\Vert}$, of
the photospheric magnetic field so that the structure function can be defined
as:
\begin{equation} 
S_q(r) = \langle | {B_{\Vert} }({\bf x} + {\bf r}) - {B_{\Vert}}({\bf x})|^q
\rangle,
\label{Sq} \end{equation}
where ${\bf x}$ is a current pixel on a magnetogram, ${\bf r}$ is the
separation vector and $q$ is the order of a statistical moment, which 
takes on real values. Angular brackets denote averaging over the magnetogram.

As we mentioned earlier, ratio of the fourth statistical moment of the
structure function to the square of the second statistical moment determines the
flatness function. However, in the case of intermittency analysis Frisch (1995)
suggested to use even higher statistical moments and to calculate the
(hyper-)flatness, namely, ratio of the sixth moment to the cube of the second
moment:
\begin{equation} 
F(r)=S_6(r)/(S_2(r))^3 \sim r^{-\kappa}.
\label{Fr} 
\end{equation} 
For non-intermittent structures, the flatness does not depend on the spatial
scale, $r$. On the contrary, for an intermittent structure, the flatness grows
as a power-law when the scale $r$ decreases (Frisch 1995, Abramenko 2005b). The
intermittency index, $\kappa$, determined as the slope of the flatness function
within a spatial range of linearity,  $\Delta r$ (Figure \ref{Fig_3}), increases
when intermittency is higher.

We applied the above technique to analyze intermittency in both spatial and
temporal domains. To process 1D time series, we modified our 2D flatness
function code based on Eqs.(\ref{Sq} - \ref{Fr}) by substituting spatial
scale by time scale, $\tau$, and magnetic field, $B_{\Vert}$, by time
series of coronal measurements.  

First, we determined the flatness functions and intermittency indices for all
selected SOT/FG magnetograms. To avoid possible contamination of the results
from the saturation effect inside the main sunspot and the vast area of
weak fields around it, only the fields enclosed by the box in Figure \ref{Fig_1}
(fast rotating spot and flare site) were taken for calculations.
The range $\Delta r = (2.3 - 23)$ Mm, as it is show in Figure \ref{Fig_3}, was
taken the same for all magnetograms.

\begin{figure}[!h] \centerline {\epsfxsize=6.0truein 
\epsffile{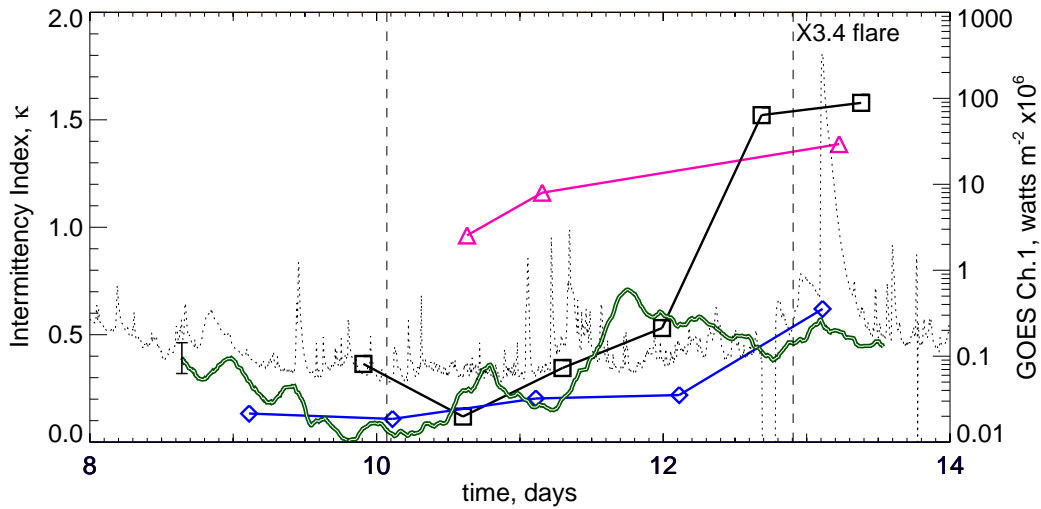}}
\caption{\sf Time variations of the intermittency index, $\kappa$, determined
from i) photospheric magnetograms ({\it double green line}, the data points were
smoothed by 35 point box car
averaging, and a typical error bar of smoothing is shown), ii) XRT data ({\it
purple triangles}), iii) Nobeyama radio data ({\it blue diamonds}), and iv) GOES
flux ({\it black squares}). For the last three cases, error bars are less than
the symbol size. 
Other notations are the same as in Figure \ref{Fig_2}.}
\label{Fig_4}
\end{figure}

Double green curve in Figure \ref{Fig_4} shows time profile of $\kappa$ in the
photosphere. The intermittency index peaked on Dec 11 at about 18:00UT (day
11.75), which is approximately 1.3 day before the X3.4 flare. This peak is
located inside the projection-free time interval (between two vertical dashed
line segments in Figure \ref{Fig_4}) and its magnitude significantly exceeds the
error bar, which allows us to consider it as a real change in complexity of the
photospheric magnetic field.

Second, to explore properties of intermittency in the solar chromosphere and
corona, we utilized time series obtained from various instruments such as
Hinode/XRT, GOES and Nobeyama radio-polarimeter. Both hard X-ray and radio
fluxes are direct traces of electrons accelerated in reconnection events.
Intermittency analysis based on these data may reveal information on the
chromospheric and coronal reconnection dynamics, i.e., reorganization in the
magnetic field.

\begin{figure}[!h] \centerline {\epsfxsize=5.0truein 
\epsffile{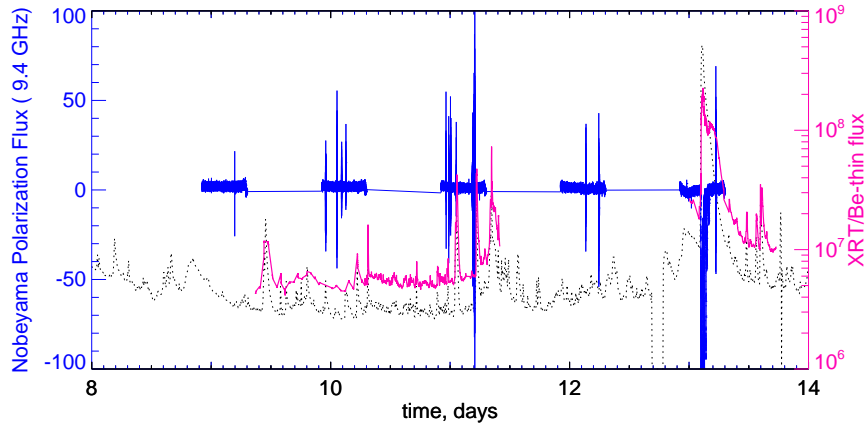}}
\caption{\sf {\it Purple - }  Time variations of the XRT/Hinode Be-thin flux
integrated over the active region area. {\it Blue - } time variations of the
radio polarization flux at 9.4 GHz from Nobeyama radioheliograph. {\it Dotted }
line represents GOES flux.} \label{Fig_5}
\end{figure}

\begin{figure}[!h] \centerline {
\epsfxsize=3.2truein \epsffile{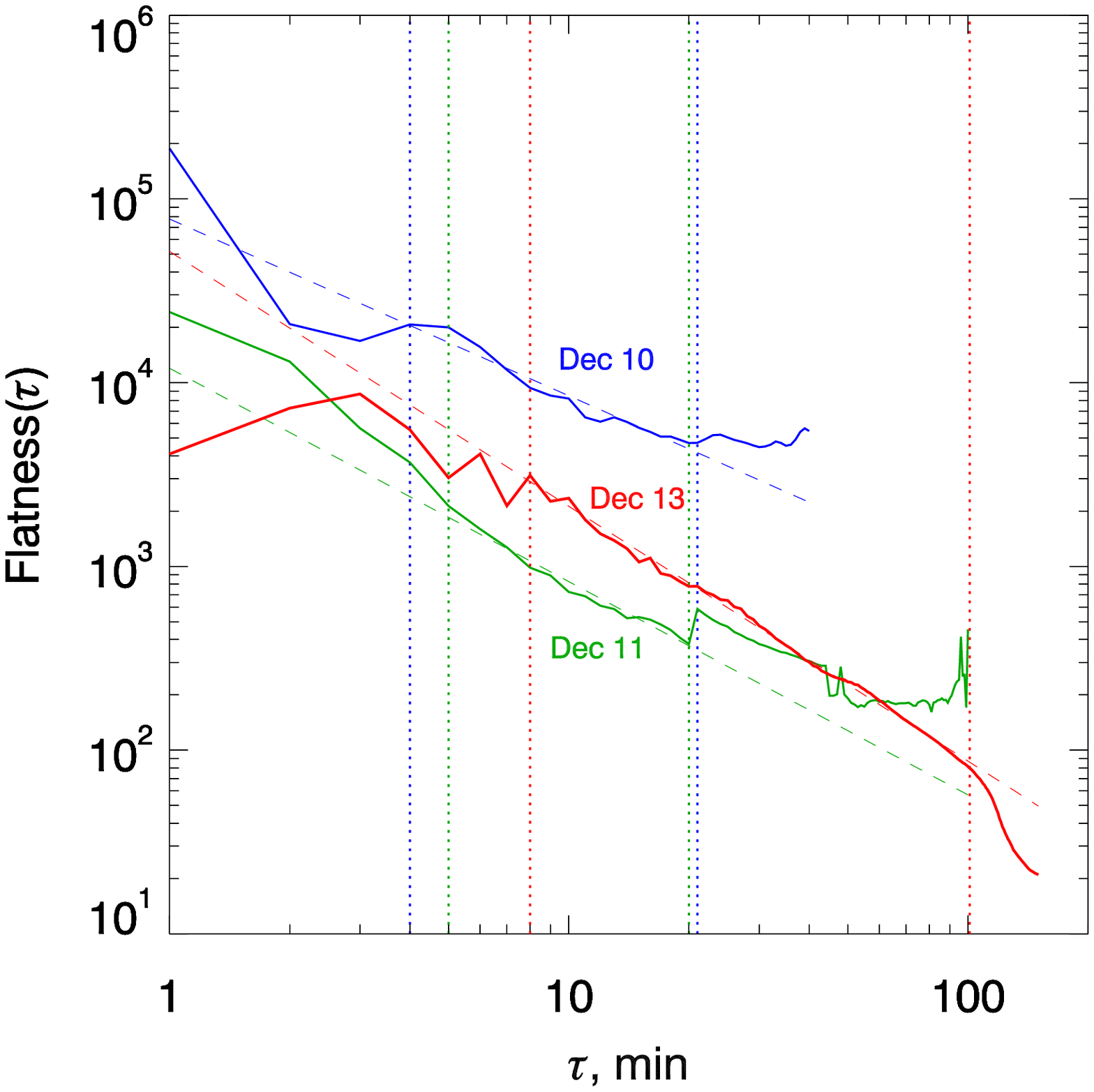}
\epsfxsize=3.2truein \epsffile{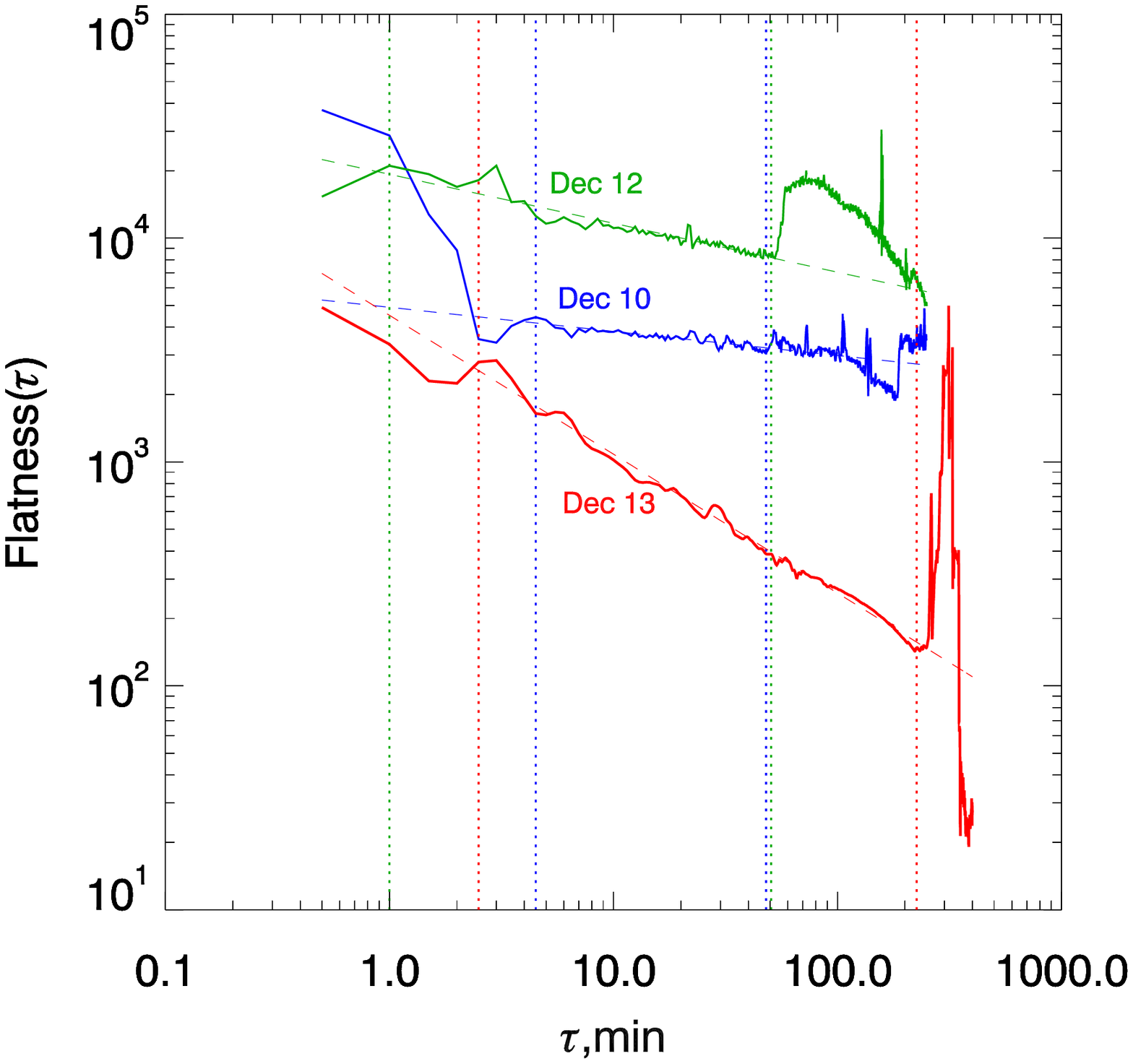}}
\caption{\sf {\it Left- } Flatness functions calculated for three time series
of XRT/Be-thin flux marked as Dec 10, Dec 11, and Dec 13. For the set Dec 10,
$\kappa=0.92 \pm 0.05$ inside $\Delta r=(4-21)$ min; for the set Dec 11, 
$\kappa=1.16 \pm 0.05$ inside $\Delta r=(5-20)$ min; for the set Dec 13,
$\kappa=1.39 \pm 0.008$ inside $\Delta r=(8-100)$ min.
{\it Right - } Flatness
functions calculated for three time series of Nobeyama polarization flux
measured at 9.4 GHz. For the set Dec 10, 
$\kappa=0.11 \pm 0.004$ inside $\Delta r=(4.3-46)$ min; for the set Dec 12,
$\kappa=0.22 \pm 0.006$ inside $\Delta r=(1-50)$ min; for the set Dec 13,
$\kappa=0.62 \pm 0.004$ inside $\Delta r=(2.5-225)$ min.}. \label{Fig_6}
\end{figure}

We used 1 min Be-thin filter data taken with XRT/Hinode instrument (Kosugi et
al. 2007, Golub et al. 2007). The XRT images were processed with the standard
SSWIDL XRT software package and the XRT flux was calculated by integrating pixel
intensities over the active region (Figure \ref{Fig_5}). The XRT data for this
time interval are not continuous. We analyzed three sub-sets acquired on 9.4 -
10.4 day, 10.4-11.4 day, and 13.0 - 13.7 day. We denoted them as Dec 10, Dec 11,
and Dec 13 data sets. Each subset contains about 1000 data points. For each
sub-set, we calculated the flatness functions (Figure \ref{Fig_6}, {\it left})
and the intermittency index $\kappa$.

The data presented in Figure \ref{Fig_6} allow to give an
explanation how we select a linear interval, $\Delta r$, for the time series. In
each case, a linear range was detected in the middle part of a spectrum. We then
extended this range in both directions and recalculated the linear fit and the
slope, $\kappa$. We continued to do so while deviations of $\kappa$ remained
inside the standard deviation of the linear fit, which was in the most of the
cases less than 0.05.

Polarization flux at 9.4 GHz from Nobeyama radioheliograph is shown in Figure
\ref{Fig_5}. The radio emission at this frequency is predominantly determined by
the gyro-resonance process and is largely controlled by the strength and
dynamics of the magnetic field above active regions (Kundu 1965; Aschwanden
2002). We re-binned these 1-second data, so that new time sampling was 30
seconds, and calculated the flatness function for each observing day thus
obtaining 5 estimations of $F(\tau)$. Three of them are presented in Figure
\ref{Fig_6} ({\it right}). The corresponding values of $\kappa$ are shown in
Figure \ref{Fig_4}.

As evidenced from both XRT and Nobeyama data (Figure \ref{Fig_6}), the slope of
the flatness function gradually steepened from Dec 10 to Dec 13, implying
a gradual increase of intermittency. One more interesting detail can be noted
in behavior of $F(\tau)$. Namely, the large-scale end of the linear interval
(i.e., interval of scales involved into intermittent process) shifts toward the
larger scales as $\kappa$ increases (see the functions for Dec 13). It means
that growing intermittency (complexity) involves increasingly larger time
scales. Note that similar behavior (involvement of larger spatial scales) we
also observed in the case of the photospheric magnetic field.

The same intermittency calculation routines were applied to the six one-day time
series of 1-minute GOES 1-8 \AA~ flux.

In Figure \ref{Fig_4} we compare the time variations of the intermittency
indices in the photosphere, chromosphere and corona. Photospheric data show an
undulating behavior with a prominent peak on Dec 11 and a gradual decrease after
that, while coronal and chromospheric indices continue to increase through Dec
13. Thus the data seem to suggest that intermittency may first increase in the
photosphere and then propagates toward the chromosphere and corona.

What processes in the photosphere and beneath are responsible for this gain of
complexity and intermittency? It is thought that convective and turbulent
motions of the magnetized plasma in photospheric and sub-photospheric layers
could be responsible for the increasing complexity (see, e.g., review by
Klimchuk 2006). Statistical comparisons (e.g., Abramenko et al., 2006;
Jing et al., 2006; Tan et al., 2007) seem to agree with this assumption. If so,
photospheric dynamics could be compared to the intermittency indices, and the
attention should be focused on the kinetics of the photospheric magnetic flux
tubes.

\section {\bf Photospheric Kinetic Vorticity}

Hinode SOT/FG level0 magnetograms are very well suitable for analyzing
horizontal displacements of magnetic elements. To measure horizontal
displacements of magnetic elements and their speed, we utilized the local
correlation tracker (LCT) technique (Strous et al. 1996), which was applied to
the same set of SOT/FG magnetograms that we used for intermittency analysis.

\begin{figure}[!h] \centerline{\epsfxsize=6.0truein
\epsffile{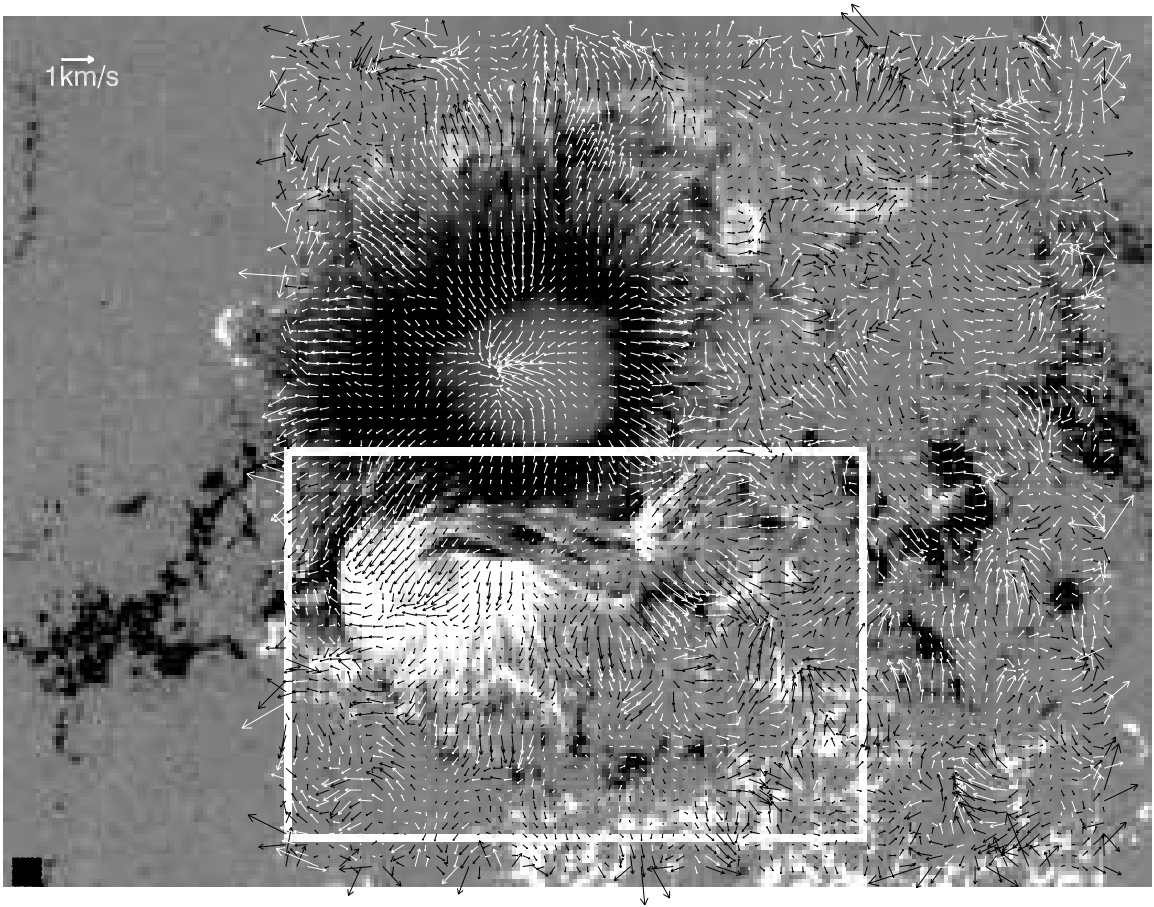}}
\caption {Horizontal velocities of magnetic elements derived by the LCT
technique for the moment 2006 Dec 12/00:48 UT. Notations are the same as in
Figure \ref{Fig_1}.}
\label{Fig_7}
\end{figure}

The FWHM of the Gaussian tracking window was $9 \times 9$ arcsec. This window
size was chosen to be an optimum trade off between the noise signal and the
spatial resolution of the flow map. A flow map was calculated for each par of
successive images in the data set and is based upon 4 min correlation interval.
Our estimation is that the solar noise (an error signal introduced by the
evolution of solar features) is less than 30 m s$^{-1}$. An example of a flow
map derived for the magnetogram in Figure \ref{Fig_1} ({\it left}) is
shown in Figure \ref{Fig_7}.

For each flow map we then calculated kinetic vorticity, $\omega$, by using
the integral formula:
\begin{equation} 
\omega(\bf r)=lim{_{s\to0} } \frac{1}{s} \int_L \bf v_{\perp}({\bf r}) dl, \label{omega} 
\end{equation}  
where the integration is performed along the contour $L$ enclosing area $s$
that contains a current point $\bf r$. In comparison with the traditional
differential technique, our approach appears to be more accurate and offers a
possibility to integrate using accurate integration methods, such as Simpson
formula \noindent ({\tt http://en.wikipedia.org/wiki/Simpson's\_rule}).

In our code, the area $s$ was represented by the Gaussian tracking window,
$s=\Delta x \times \Delta y$. The components of the transverse velocity,
$v_x(i,j)$ and $v_y(i,j)$, were interpolated into a refined mesh of 
$\Delta x/2 \times \Delta y/2$ pixel size, and then the integration in 
the CCW direction along the contour  $L=[\Delta x,\Delta y, -\Delta x,
-\Delta y]$, which encloses a current point ${\bf r}(i,j)$, was performed.
Integrals along each side of $L$ were calculated by the Simpson's formula. For
example, an integral along the positive $x$-direction (the bottom side of $L$)
was 
\begin{equation}
I_1= \frac{\Delta x}{6} [v_x(m,n)+4v_x(m+1,n)+v_x(m+2,n)].
\end{equation}
Here indices $m,n$ belong to the refined mesh, so that $m=2i$ and $n=2j$. The
sum of four integrals divided by $s$ gives us the estimation of the kinetic
vorticity at a current point ${\bf r}(i,j)$.

We would like to mention that an analog to the kinetic vorticity is electric
current which can be calculated by constituting the flow field $v_{\perp}$ in
Eq. (\ref{omega}) with the horizontal magnetic field $B_{\perp}$. Having the map
of $\omega(\bf r)$, we calculated the averaged over the area squared kinetic
vorticity,  $<\omega^2>$.  This parameters characterizes the dissipation rate of
the kinetic energy in the photosphere caused via random motions of footpoints of
magnetic flux tubes.

\begin{figure}[!h] \centerline {\epsfxsize=5.5truein 
\epsffile{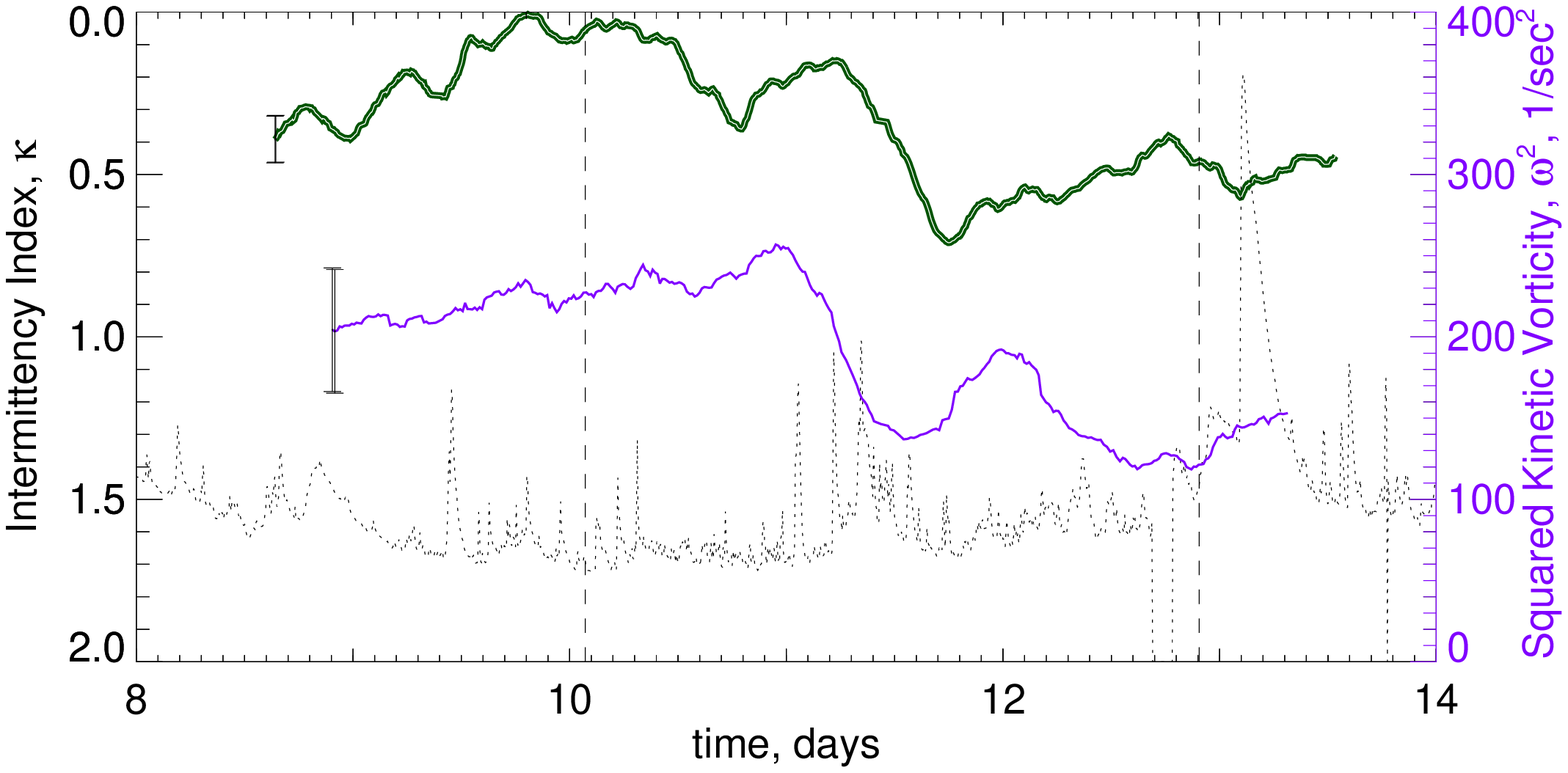}}
\caption{\sf {\it Purple - } Time variations of the squared kinetic vorticity,
$<\omega^2>$, averaged over area. {\it Green - } time variations of the
intermittency index, $\kappa$, in the photosphere. To ease the comparison, left
axis was reversed. {\it Dotted} line is GOES flux. Other notations are the same
as in Figure \ref{Fig_2}.}
\label{Fig_8}
\end{figure} 

In Figure \ref{Fig_8} we compared the time variation of $<\omega^2>$ and the
photospheric intermittency index, $\kappa$. The plot shows that there exists
a systematic time lag between the two curves with the intermittency being
delayed. Cross-correlation analysis showed that the delay is approximately 4
hours. This indicates that the gain of intermittency in the photosphere is
preceded by the enhanced rate of kinetic energy dissipation. In other words, the
increase in the kinetic vorticity (or self-rotation of plasma structures) leads
to subsequent increase of complexity of the photospheric magnetic fields.

\section { Discussion and Conclusion }

Presented here case study is devoted to the analysis of emergence of a rotating
sunspot in the close vicinity of a mature spot of opposite polarity. The fact of
the rotation was reported by Nightingale et al. (2007) and is considered as a
possible energy source for the X-class flares in this active region. This case
demonstrates a rather typical evolution of a delta structure: a highly twisted
and stressed magnetic flux rope emerges nearby the pre-existing sunspot. The
emerging sunspot of opposite polarity ``screws'' into the active region magnetic
environment. The interaction of the new and old magnetic flux is accompanied by
a chain of processes: interchange reconnection at the interaction boundary,
propagation of the magnetic stress and helicity into the corona and gain of
complexity in chromospheric and coronal magnetic fields. As a result of magnetic
reconnections on a variety of spatial scales, new magnetic connections form.
Thus, in this particular case of NOAA AR 10930, Kubo et al. (2007) reported that
one day before the X3.4 flare, Ca H II bright loops began appear near the
polarity inversion line. In general, continuous injection of energy and
associated magnetic re-arrangements may increase magnetic energy stored in
active region. The result often can be a X-class flare, associated with a
coronal mass ejection. In NOAA AR 10930 two powerful X-ray flares were observed
on Dec 13 and 14, 2006.

What can be added to this scenario from the present research?

The emergence of the rotating sunspot was associated with undulating variations
of the squared kinetic vorticity, $<\omega^2>$. The most pronounced peak
observed about 2 days before the X3.4 flare was followed by an abrupt fall off.
Approximately 4 hours after this enhanced activity of photospheric plasma
vorticies we observed a peak of the intermittency index of the photospheric
magnetic field. Intermittency can be considered as a measure of complexity of
the field and implies the tendency of the field to concentrate into extremely
strong, widely separated flux tubes (or sheets) and a burst-like behavior of
energy release in time. Intermittency can increase, in particular, due to
fragmentation and merging, as well as due to abrupt intrusion of strong
entities. We therefore suggest that the abrupt exhausting of strong plasma
vorticies presented more freedom for magnetic flux tubes and thus facilitated
fragmentation and merging processes, eventually resulting in the gain of
intermittency in the photosphere. Strongly intermittent photospheric magnetic
field represents a more stressed magnetic configuration permeated by a multitude
of magnetic field discontinuities, which tend to propagate upward due to the
magnetic tension.

After the photospheric magnetic intermittency peaked, the chromospheric and
coronal intermittency continued to increase during, at least, one more day.
Approximately 1.3 days after the peak of photospheric intermittency, the first
X-class flare was launched. The data allow us to suggest that the magnetic field
first became highly intermittent in the photosphere, and then intermittency had
penetrated toward the corona, either due to diffusion of magnetic
discontinuities, or due to waves of various types and their interactions. As a
result, a highly critical state of the coronal magnetic configuration was
reached, the state which at any instant, due to any perturbance, may lead to an
eruption. 

In the framework of the intermittency concept, the phenomenological scenario for
development of a magnetic structure can be regarded as evolution of a non-linear
dynamical dissipative system, thus offering a more general insight into the
phenomenon. Indeed, a non-linear dynamical dissipative system is believed to
evolve toward its attractor, which is a self-organized criticality (SOC) state.
The SOC state is characterized by non-rare occurrence of extremely large
fluctuations. Which, in turn, results in non-Gaussian distributions of various
parameters (recall power-law distributions of flare energy, flare duration,
etc., see, e.g.,  Lu \& Hamilton 1991; Charbonneau et al. 2001) and highly
intermittent, or, in other words, multifractal organization of the system, as in
temporal, so in spatial domain. At the SOC state, any perturbance can
provoke an eruption of any size, and thus an eruption cannot be predicted in
advance.

Of cause, a {\it short-time} prediction may be quite possible (recall that a
snow avalanche, for example, can be "predicted" several seconds in advance by
sound), however, this is actually a post factum prediction based on a finite
speed of avalanche propagation. The same can be said about hard X-ray
precursors of H$\alpha$ flares. 

Recently, Leka \& Barnes (2007) have come to the conclusion that an individual
snapshot of an active region hardly bears information about the time of oncoming
flare. In our opinion, this inference perfectly agrees with the concept of
non-linear dynamical dissipative system evolution: eruptions cannot be
predicted. One can only say that {\it when} the system reached the SOC state,
strong eruptions can happen frequently enough, along with a multitude of smaller
ones. One of the ways to make a step ahead is to analyze whether the system
reached the SOC state or it is still at the stage of accumulating the energy and
complexity? How the SOC state can be reached in the corona and what is the role
of the photosphere? In the present study we undertook an attempt tackle these
questions.
One should keep in mind also that the SOC concept, as any other theory, has
their advantages and disadvantages being under continuous elaboration,
see, e.g., Belanger et al. (2007).

As to the usefulness of the intermittency/multifractality concept for
understanding of solar phenomena, it is worth to mention the long-standing
problem of appearance of low plasma conductivity in the corona, especially
during a flare. To explain a solar eruption on the scale of an active region
that can last of about 100 minutes, it is necessary to imply the presence of
super strong electric currents $(\sim 10^{10} A/km^2)$ inside a very thin layers
($<100\ m$, Priest 1982). A fractal concept of coronal magnetic fields can easy
meet these requirements. Indeed, a self-similar fractal allows existence of
super thin branches (magnetic sheets or tubes) whereas a percolation state,
i.e., a large-scale avalanche of the SOC state, implies formation of super
strong currents at singular branches of the cluster.

The authors would like to thank the anonymous referee for stimulating
discussions and suggestions for improvement of the text.
This work was supported, in part, by NASA NNG05GN34G, NASA NNX07AT16G grants,
NSF grant ATM-0716512. Hinode is a Japanese mission developed and launched by
ISAS/JAXA, with NAOJ as domestic partner and NASA and STFC (UK) as international
partners. It is operated by these agencies in co-operation with ESA and  NSC
(Norway). 

{}
\end{document}